\documentclass[a4paper, 10 pt, conference]{ieeeconf}  % Comment this line out if you need a4paper
\usepackage[utf8]{inputenc}
\usepackage[T1]{fontenc}

\IEEEoverridecommandlockouts                              % This command is only needed if                                                          
\usepackage{graphics} % for pdf, bitmapped graphics files
\usepackage{amsmath}
\usepackage{amssymb}  
\usepackage{color}
\usepackage{units}
\usepackage{tikz}
\usepackage{booktabs}
\usepackage{multirow}
\usepackage{array}
\usepackage{caption}
\usepackage{blindtext}
\usepackage{tkz-kiviat}
\usepackage{pgfplotstable}
\usetikzlibrary{arrows}
\pgfplotsset{compat=newest}
\usepackage{xcolor}
\usepackage{subcaption}
\usepackage[nolist]{acronym}

\newcommand{\sig}{\mathbf{\hat{\sigma}}}

\newcommand{\rhat}{\hat{r}}

\newcommand{\z}{\mathbf{z}}

\newcommand{\norm}[1]{\left\lVert#1\right\rVert}

\title{\LARGE \bf
Uncertainty depth estimation with gated images for 3D reconstruction}

\author{Stefanie Walz$^{1,2}$, Tobias Gruber$^{1,2}$, Werner Ritter$^{1}$, Klaus Dietmayer$^{2}$
\thanks{$^{1}$ The authors are with Mercedes-Benz AG, Wilhelm-Runge-Str. 11, 89081 Ulm, Germany. E-mail:  {\tt\scriptsize stefanie.walz@daimler.com}, {\tt\scriptsize tobias.gruber@daimler.com}, {\tt\scriptsize werner.r.ritter@daimler.com}}
\thanks{$^{2}$ The authors are with the Institute of Measurement, Control and Microtechnology, Ulm University, Albert-Einstein-Allee 41, 89081 Ulm, Germany. E-mail: {\tt\scriptsize klaus.dietmayer@uni-ulm.de}}
}

\begin{document}

% taken from https://designnavigator.daimler.com/Daimler_Color_System

\definecolor{dai_ligth_grey}{RGB}{230,230,230}
\definecolor{dai_ligth_grey20K}{RGB}{200,200,200}
\definecolor{dai_ligth_grey40K}{RGB}{158,158,158}
\definecolor{dai_ligth_grey60K}{RGB}{112,112,112}
\definecolor{dai_ligth_grey80K}{RGB}{68,68,68}

\definecolor{dai_petrol}{RGB}{0,103,127}
\definecolor{dai_petrol20K}{RGB}{0,86,106}
\definecolor{dai_petrol40K}{RGB}{0,67,85}
\definecolor{dai_petrol80}{RGB}{0,122,147}
\definecolor{dai_petrol60}{RGB}{80,151,171}
\definecolor{dai_petrol40}{RGB}{121,174,191}
\definecolor{dai_petrol20}{RGB}{166,202,216}

\definecolor{dai_deepred}{RGB}{113,24,12}
\definecolor{dai_deepred20K}{RGB}{90,19,10}
\definecolor{dai_deepred40K}{RGB}{68,14,7}

\definecolor{mittelblau}{RGB}{0, 126, 198}
\definecolor{violettblau}{cmyk}{0.9, 0.6, 0, 0}
\definecolor{rot}{RGB}{238, 28 35}
\definecolor{apfelgruen}{RGB}{140, 198, 62}
\definecolor{gelb}{RGB}{255, 229, 0}
\definecolor{orange}{RGB}{244, 111, 33}
\definecolor{pink}{RGB}{237, 0, 140}
\definecolor{lila}{RGB}{128, 10, 145}
\definecolor{hellgrau}{RGB}{224, 224, 224}
\definecolor{mittelgrau}{RGB}{128, 128, 128}
\definecolor{dunkelgrau}{RGB}{80,80,80}
\definecolor{anthrazit}{RGB}{19, 31, 31}

\newcommand{\networkLayer}[7]{
	\def\a{#1} % Used to distinguish input resolution for current layer.
	\def\b{0.02}
	\def\c{#2} % Width of the cube to distinguish number of input channels for current layer.
	\def\t{#3} % X offset for current layer.
	\def\d{#4} % Y offset for current layer.
	\def\e{#5} % Z offset for current layer.
	
	% Draw the layer body.
	\draw[line width=0.3mm](\c+\t,\e,\d) -- (\c+\t,\e+\a,\d) -- (\t,\e+\a,\d);                                                      % back plane
	\draw[line width=0.3mm](\t,\e,\a+\d) -- (\c+\t,\e,\a+\d) node[midway,below] {#7} -- (\c+\t,\e+\a,\a+\d) -- (\t,\e+\a,\a+\d) -- (\t,\e,\a+\d); % front plane
	\draw[line width=0.3mm](\c+\t,\e,\d) -- (\c+\t,\e,\a+\d);
	\draw[line width=0.3mm](\c+\t,\e+\a,\d) -- (\c+\t,\e+\a,\a+\d);
	\draw[line width=0.3mm](\t,\e+\a,\d) -- (\t,\e+\a,\a+\d);
	
	% Recolor visible surfaces
	\filldraw[#6] (\t+\b,\e+\b,\a+\d) -- (\c+\t-\b,\e+\b,\a+\d) -- (\c+\t-\b,\e+\a-\b,\a+\d) -- (\t+\b,\e+\a-\b,\a+\d) -- (\t+\b,\e+\b,\a+\d); % front plane
	\filldraw[#6] (\t+\b,\e+\a,\a-\b+\d) -- (\c+\t-\b,\e+\a,\a-\b+\d) -- (\c+\t-\b,\e+\a,\b+\d) -- (\t+\b,\e+\a,\b+\d);
	
	% Colored slice.
	\ifthenelse {\equal{#6} {}}
	{} % Do not draw colored slice if #4 is blank.
	{\filldraw[#6] (\c+\t,\e+\b,\a-\b+\d) -- (\c+\t,\e+\b,\b+\d) -- (\c+\t,\e+\a-\b,\b+\d) -- (\c+\t,\e+\a-\b,\a-\b+\d);} % Else, draw a colored slice.
}

\newcommand{\skipConnection}[4]{
	\draw[->,densely dashed](#1,#3) -- (#1,#4) -- (#2,#4) -- (#2,#3);
}

\newcommand{\radarChart}[1]{
	\begin{tikzpicture}
	\node at (0,0) {
		\begin{tikzpicture}[scale=1, every node/.style={scale=1}]
		
		\tkzKiviatDiagramFromFile[
		scale = 0.5,
		label distance = 0,
		gap = 0.5,
		label space = 1.5,
		step = 1,
		lattice = 10]{#1}
		
		\tkzKiviatLineFromFile%
		[thick,
		color = \monodepthColor,
		mark size = 4pt]{#1}{3}
		
		\tkzKiviatLineFromFile%
		[thick,
		color = \psmColor,
		mark size = 4pt]{#1}{1}
		
		\tkzKiviatLineFromFile%
		[thick,
		color = \sgmColor,
		mark size = 4pt]{#1}{2}
		
		\tkzKiviatLineFromFile%
		[thick,
		color = \sparseColor,
		mark size = 4pt]{#1}{4}
		
		\tkzKiviatLineFromFile%
		[thick,
		densely dashed,
		color = \gatedColor,
		mark size = 4pt]{#1}{5}
		
		\tkzKiviatLineFromFile%
		[thick,
		color = \gatedColor,
		densely dotted,
		mark size = 4pt]{#1}{6}
		
		\tkzKiviatLineFromFile%
		[thick,
		color = \gatedColor,
		mark size = 4pt]{#1}{7}

		\end{tikzpicture}
	};
	
	\node[] at (0,3.75) {
		\begin{tikzpicture}[scale=0.9, every node/.style={scale=0.9}]
		\begin{axis}[%
		hide axis,
		xmin=10,
		xmax=10,
		ymin=0,
		ymax=0.4,
		legend style={draw=white!15!black,legend cell align=left},
		legend style={legend cell align=left, legend},
		legend columns=3,
		]
		\addlegendimage{\gatedColor, thick, densely dashed}
		\addlegendentry{\gatedLegend~\cite{Gruber2019}};
		\addlegendimage{\gatedColor, thick, densely dotted}
		\addlegendentry{G2D+ Uncertainty};
		\addlegendimage{\gatedColor, thick}
		\addlegendentry{G2D+ Filter (80\,\%)};
		\addlegendimage{\monodepthColor, thick}
		\addlegendentry{\monodepthLegend};
		\addlegendimage{\psmColor, thick}
		\addlegendentry{\psmLegend};
		\addlegendimage{\sgmColor, thick}
		\addlegendentry{\sgmLegend};
		\addlegendimage{\sparseColor, thick}
		\addlegendentry{\sparseLegend};

		\end{axis}
		\end{tikzpicture}};
\end{tikzpicture}
}

\newcommand{\psmColor}{orange}
\newcommand{\sgmColor}{lila}
\newcommand{\sparseColor}{mittelblau}
\newcommand{\monodepthColor}{rot}
\newcommand{\gatedColor}{apfelgruen}

\newcommand{\psmLegend}{PSMnet \cite{Chang2018}}
\newcommand{\sgmLegend}{SGM \cite{hirschmuller2005accurate}}
\newcommand{\sparseLegend}{Sparse-to-Dense \cite{ma2018self}}
\newcommand{\monodepthLegend}{Monodepth \cite{Godard2017}}
\newcommand{\gatedLegend}{Gated2Depth}

\begin{acronym}
 \acro{SNR}{signal-to-noise ratio}
 \acro{SfM}{structure from motion}
 \acro{RIP}{range intensity profile}
 \acro{ToF}{time-of-flight}
 \acro{BNN}{Bayesian neural network}
 \acro{VSCEL}{vertical-cavity surface-emitting laser}
 \acro{GTA V}{Grand Theft Auto V}
 \acro{KL}{Kullback-Leibler}
 \acro{ELBO}{evidence lower bound}
 \acro{MAE}{mean absolute error}
 \acro{DC}{depth completion}
 \acro{LiDAR}{light detection and ranging}
 \acro{GPU}{graphics processing unit}
 \acro{RMSE}{root-mean-squared error}
 \acro{SIlog}{scale invariant logarithmic error}
\end{acronym}

\maketitle
\thispagestyle{empty}
\pagestyle{empty}

%%%%%%%%%%%%%%%%%%%%%%%%%%%%%%%%%%%%%%%%%%%%%%%%%%%%%%%%%%%%%%%%%%%%%%%%%%%%%%%%
\begin{abstract}
Gated imaging is an emerging sensor technology for self-driving cars that provides high-contrast images even under adverse weather influence. 
It has been shown that this technology can even generate high-fidelity dense depth maps with accuracy comparable to scanning LiDAR systems. 
In this work, we extend the recent \emph{Gated2Depth} framework with aleatoric uncertainty providing an additional confidence measure for the depth estimates. 
This confidence can help to filter out uncertain estimations in regions without any illumination.
Moreover, we show that training on dense depth maps generated by LiDAR depth completion algorithms can further improve the performance.
\end{abstract}

%%%%%%%%%%%%%%%%%%%%%%%%%%%%%%%%%%%%%%%%%%%%%%%%%%%%%%%%%%%%%%%%%%%%%%%%%%%%%%%%
\section{INTRODUCTION} 

Self-driving vehicles require a very detailed perception of their environment in order to move around safely. 
While 3D information is crucial to understand the scenery and to detect free space, information about texture enables to classify objects and predict their interaction.
The environment can be perceived by a variety of sensors, each of them with benefits and drawbacks.
Cameras are low-priced and provide high-resolution images that are essential for capturing textures.
However, 3D scene reconstruction from a single image is an ill-posed problem \cite{smolyanskiy2018importance} and stereo setups are limited in range resolution \cite{davies2004machine}.
In contrast, laser scanners offer very accurate depth measurements, but at very low spatial resolution and for a high cost \cite{chan2008noise}.
In addition to 3D information, radar systems can measure the velocity of objects.
Despite of the huge progress in radar development in recent years, radars are still low resolution and can capture neither texture nor fine 3D structures.
While radars still perform reasonable in adverse weather conditions, laser scanners completely fail in scattering environments \cite{bijelic2018benchmark} and standard cameras suffer from strong contrast degeneration \cite{bijelic2018benchmarking}.

\begin{figure}
\centering
\hspace*{-1mm}
\begin{tikzpicture}
\footnotesize
\node[anchor=west,inner sep=0pt] (slice1) at (0,0) {\includegraphics[width=0.328\columnwidth]{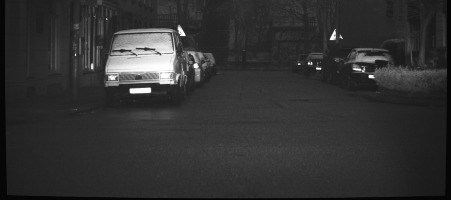}};
\node[inner sep=0pt] (slice2) at (4.35,0) {\includegraphics[width=0.328\columnwidth]{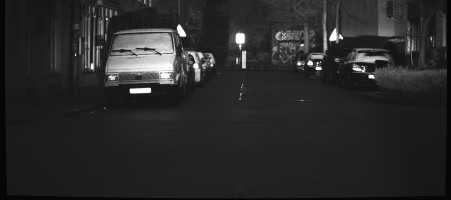}};
\node[anchor=west,inner sep=0pt] (slice3) at (5.8,0) {\includegraphics[width=0.328\columnwidth]{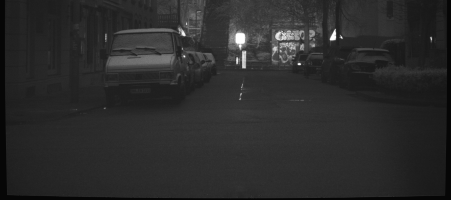}};

\node[inner sep=0pt] (depth) at (2.14,-4.0) {\includegraphics[width=0.495\columnwidth]{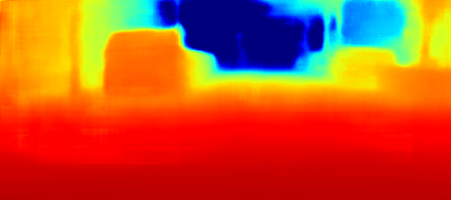}};

\node[inner sep=0pt] (uncertainty) at (6.5,-4.0) {\includegraphics[width=0.495\columnwidth]{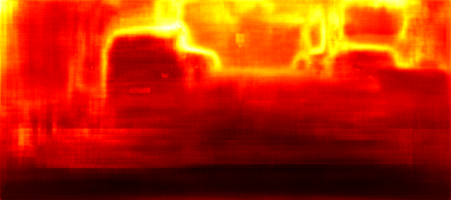}};

\node[anchor=west,inner sep=0pt,white] at (4.46,-4.75) {Uncertainty};
\node[anchor=west,inner sep=0pt,white] at (0.1,-4.75) {Depth map};

\node[anchor=west,inner sep=0pt,white] at (0.1,-0.45) {Slice 1};
\node[anchor=west,inner sep=0pt,white] at (3.0,-0.45) {Slice 2};
\node[anchor=west,inner sep=0pt,white] at (5.9,-0.45) {Slice 3};

\node[text width=15mm, align=center] at (1.2,-1.8) {\emph{Gated2Depth} network};

\node[draw,thick,dai_deepred,rounded corners=.1cm,inner sep=5pt, align=center,minimum height=15mm, text width=40mm] (network) at (4.35,-1.8) {};

\node at (4.35,-1.8) {
	\begin{tikzpicture}[scale=0.2, every node/.style={scale=0.2}]
	% ENCODER
	\networkLayer{4.0}{0.1}{0.0}{0.0}{0.0}{color=dai_ligth_grey40K}{}    % S1
	\networkLayer{4.0}{0.1}{0.4}{0.0}{0.0}{color=dai_ligth_grey40K}{}      % S2
	\networkLayer{2.0}{0.2}{0.8}{0.0}{0.0}{color=dai_petrol}{}    % S1
	\networkLayer{2.0}{0.2}{1.3}{0.0}{0.0}{color=dai_ligth_grey40K}{}        % S2
	\networkLayer{1.0}{0.4}{1.8}{0.0}{0.0}{color=dai_petrol}{}    % S1
	\networkLayer{1.0}{0.4}{2.5}{0.0}{0.0}{color=dai_ligth_grey40K}{}        % S2
	\networkLayer{0.5}{0.8}{3.2}{0.0}{0.0}{color=dai_petrol}{}    % S1
	\networkLayer{0.5}{0.8}{4.3}{0.0}{0.0}{color=dai_ligth_grey40K}{}        % S2
	\networkLayer{0.25}{1.6}{5.4}{0.0}{0.0}{color=dai_petrol}{}    % S1
	\networkLayer{0.25}{1.6}{7.3}{0.0}{0.0}{color=dai_ligth_grey40K}{}        % S2
	
	% DECODER
	\networkLayer{0.5}{0.8}{9.3}{0.0}{0.0}{color=dai_deepred}{} % S1
	\networkLayer{0.5}{0.8}{10.4}{0.0}{0.0}{color=dai_ligth_grey40K}{}       % S2
	\networkLayer{1.0}{0.4}{11.7}{0.0}{0.0}{color=dai_deepred}{}       % S1
	\networkLayer{1.0}{0.4}{12.4}{0.0}{0.0}{color=dai_ligth_grey40K}{}       % S2
	\networkLayer{2.0}{0.2}{13.6}{0.0}{0.0}{color=dai_deepred}{}       % S1
	\networkLayer{2.0}{0.2}{14.1}{0.0}{0.0}{color=dai_ligth_grey40K}{}       % S2
	\networkLayer{4.0}{0.1}{15.4}{0.0}{0.0}{color=dai_deepred}{}       % S1
	\networkLayer{4.0}{0.1}{15.8}{0.0}{0.0}{color=dai_ligth_grey40K}{}       % S2
	
	% SKIP CONNECTIONS
	\skipConnection{4.6}{9.6}{0.6}{1.1}
	\skipConnection{2.7}{11.9}{1.1}{1.7}
	\skipConnection{1.4}{13.7}{2.1}{2.4}
	\draw[->,densely dashed](0.6,3.2) -- (14.5,3.2);
	\end{tikzpicture}
};	

\draw[->] (slice1.south) |- (3.85,-0.8) -| ([xshift=-5mm]network.north);
\draw[->] (slice2.south) |- (4.35,-0.8) -| (network.north);
\draw[->] (slice3.south) |- (4.85,-0.8) -| ([xshift=5mm]network.north);

\draw[->] ([xshift=-2.5mm]network.south) |- (3.5,-2.8) -| (depth.north);
\draw[->] ([xshift=2.5mm]network.south) |- (5,-2.8) -| (uncertainty.north);

\node[inner sep=0pt] at (2.14,-5.25) {
\includegraphics[width=0.475\columnwidth]{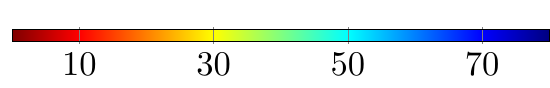}
};

\node[inner sep=0pt] at (6.5,-5.25) {
\includegraphics[width=0.475\columnwidth]{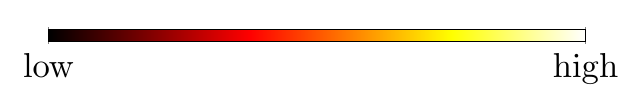}
};

\end{tikzpicture}
\caption{We extend the recent \emph{Gated2Depth} architecture \cite{Gruber2019} with an aleatoric uncertainty framework that converts three gated images into a high-resolution depth map and an uncertainty measure.}
\label{fig:teaser}
\vspace*{-7mm}
\end{figure}
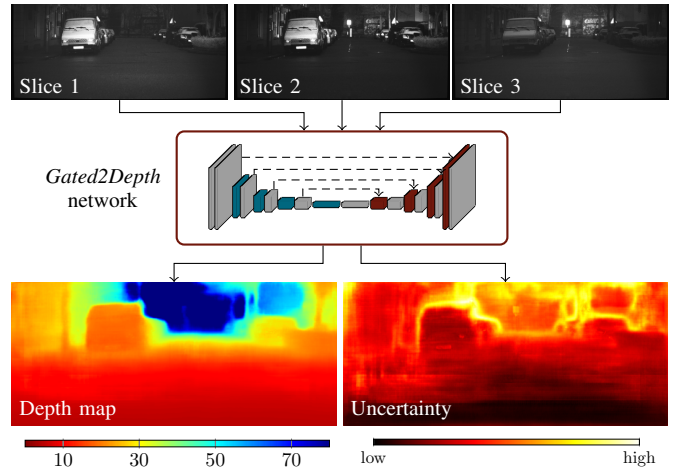

For safe autonomous driving in any conditions, all of these sensors are probably required because there is no single sensor that can solve everything reliably and sensors fail asymmetrically under adverse weather \cite{bijelic2018benchmark,bijelic2018benchmarking,bijelic2020seeing}.
Therefore, the fusion of multiple sensors is the key for reliable and accurate environment perception.
For sensor fusion, it is extremely helpful if every sensor does not only deliver measurements but also a measure about the quality or uncertainty of these measurements.
Many stereo algorithms \cite{scharstein2002taxonomy,hirschmuller2005accurate} for example provide a confidence metric that rates the provided disparity measurement.
Sensor stream uncertainties help the fusion algorithm to interpret and weight the incoming sensor measurements.
As an alternative, information content can be classified by calculating the entropy \cite{bijelic2020seeing}.

Gated vision has been presented as an active imaging solution that significantly reduces backscatter and provides high-contrast images even in fog and rain \cite{bijelic2018benchmarking}.
Additionally, the scene geometry can be reconstructed at image resolution by processing at least two images with different delays \cite{Laurenzis2007}.
As gated imaging is an active system, less illuminated areas with low \ac{SNR} offer less accuracy of 3D sensing \cite{gohler2012range}.
Generative models have been introduced in \cite{Gruber2019} in order to improve depth estimation in these areas by incorporating context information.
Nevertheless, the confidence of the additionally generated information is not yet evaluated and is obviously not constant.
For a sensor fusion algorithm, it would be beneficial to know for each pixel if the depth is measured, or generated by context and experience, and how certain the generative model is about its prediction.

In recent years, uncertainty estimation of neural networks based on Bayesian networks \cite{gal2015bayesian} has found its way into a wide field of applications such as scene segmentation~\cite{kendall2017uncertainties} and object detection~\cite{kraus2019uncertainty,choi2019gaussian}.
In this work, we extend the generative model for gated depth estimation \cite{Gruber2019} with additional confidence estimation for better interpretable depth estimates.
We show that the estimated depth confidence provides much higher quality than using the \ac{SNR} of the input images.
By filtering out a very small number of extreme outliers with very low confidence, we significantly improve the overall performance of 3D reconstruction.

%%%%%%%%%%%%%%%%%%%%%%%%%%%%%%%%%%%%%%%%%%%%%%%%%%%%%%%%%%%%%%%%%%%%%%%%%%%%%%%%
\section{RELATED WORK} 

\paragraph{3D sensing} 
3D reconstruction from images is one of the fundamental challenges in computer vision.
While \ac{SfM} \cite{ranftl2016dense,Ummenhofer2017demon} exploits the motion of the camera to obtain multiple views for depth estimation by triangulation, stereo vision \cite{scharstein2002taxonomy,hirschmuller2005accurate} or multi-view methods \cite{hartley2003multiple} rely on a fixed and calibrated camera setup.
Multi-view relies on finding correspondences and therefore suffer in texture-less regions and in case of occlusions.
There is already work on the confidence of the stereo estimation, either directly in the stereo algorithm \cite{scharstein2002taxonomy,hirschmuller2005accurate} or as a post-processing step \cite{tosi2018beyond}.
Static monocular setups instead leverage visual cues such as shading \cite{woodham1980photometric}, perspective, or relative size \cite{saxena2006learning}, in recent years usually driven by deep neural networks \cite{eigen2014depth,Laina2016,Garg2016,Godard2017,Kuznietsov2017,garg2019learning}.
However, most of the current approaches do not offer a probabilistic interpretation of the depth estimation.
Some first approaches for monocular depth prediction with confidence interpretation have been reported in \cite{yang2019bayesian, liu2019neural}.
Gated imaging is related to other active systems such as \ac{LiDAR} \cite{christian2013survey} and \ac{ToF} cameras \cite{horaud2016overview}.
It was shown in \cite{reynolds2011capturing} that for \ac{ToF} cameras it is insufficient to remove inaccurate estimates by simply thresholding low-amplitude values.
Uncertainty can be either captured by a combination of distance, amplitude, their temporal and spatial variations, or learned by a regressor such as a random forest \cite{reynolds2011capturing}.
\ac{LiDAR} depth completion combines a sparse \ac{LiDAR} point cloud with RGB images in order to generate a high resolution depth map \cite{chen2018estimating,jaritz2018sparse,ma2018sparse,ma2018self}.
A first approach where confidence maps are additionally learned during depth completion has been presented in \cite{van2019sparse}.

\paragraph{Uncertainty estimation}
Bayesian modeling offers the possibility to capture epistemic and aleatoric uncertainty in deep learning frameworks. While epistemic uncertainty represents uncertainty in the model parameters, aleatoric uncertainty depicts noise inherent in the observation. Utilizing \acp{BNN}, epistemic uncertainty can be modeled by inferring a posterior distribution over the model weights \cite{mackay1992practical}. 
This is realized by replacing deterministic weights with stochastic weights following prior distributions. 
Generally, dropout variational inference is utilized to approximate \acp{BNN} \cite{gal2016dropout}. 
That means that dropout is performed not only during training but also at test time in order to sample the posterior distribution.
This approach was applied to camera localization estimation  \cite{kendall2016modelling}, semantic segmentation \cite{kendall2015bayesian}, and open-set object detection tasks \cite{miller2018dropout}.
Aleatoric uncertainty is modeled by placing a distribution over the output of the model. 
It is utilized to improve object detection \cite{feng2019leveraging} and to adjust weights of multi-task loss functions automatically \cite{kendall2018multi}. 
Besides individual modeling of aleatoric and epistemic uncertainty,  both uncertainties can be captured simultaneously. 
Kendall \textit{et al.} \cite{kendall2017uncertainties} introduced a framework for classification and regression tasks, which can model either aleatoric or epistemic uncertainty alone or both together. 
This approach is further developed for two-stage \cite{feng2018towards} and one-stage \cite{kraus2019uncertainty,choi2019gaussian} object detection. 
In this work, we model only aleatoric uncertainty since it can be extracted without time-consuming dropout sampling enabling our proposed framework to be run as a real-time application.

\paragraph{Gated depth estimation}
Active gated imaging requires a sensitive image sensor and an illumination source. 
The synchronization of sensor gate and light source delivers the reflectivity of a scene in a certain range which enables the view through scattered environments. 
Heckman and Hodgson \cite{Heckman1967} were the first to take advantage of this visualization technique by using it to extend the range of visibility underwater. 
The active gated imaging technique offers two-dimensional images of the scene. 
To get a three-dimensional output, multiple gated images must be captured with different delays. 
Various methods have been developed for gated depth estimation, such as the time-slicing method. 
In this method, several two-dimensional gated images must be recorded in sequence where the gate delay is increased after each image. 
The resulting gate delay profiles are used to estimate the depth pixel-wise by threshold the rising or falling edge, determining the maximum, or by computing the weighted average of the profile \cite{Busck2004}. 
Additionally, Andersson \cite{Andersson2006} resolved the depth by least-squares parameter fitting and data feature positioning. 
However, high-range accuracy requires small scanning step sizes resulting in a significant rise in capture time and processing effort. 
To overcome this problem, gain modulation and super-resolution depth mapping have been introduced. 
The gain modulation method exploits a pulsed laser and an intensified camera to recover the depth information with a gain-modulated and gain-constant image. 
Gain-modulated images are generated by linearly \cite{Xiuda2008} or exponentially \cite{Jin2009} increasing the gain of the intensifier during the gated time. 
This ensures independence of the laser pulse shape. 
The super-resolution depth mapping method exploits the knowledge of the \acp{RIP} that are given by the convolution of the illuminating laser pulse and the sensor gate. 
To determine the depth, at least two gated images with distinct delay and overlapping \acp{RIP} are required. 
The use of trapezoid- \cite{Laurenzis2007} or triangular-shaped \acp{RIP} \cite{Xinwei2013} enables range estimation by exploiting the strong linear dependencies of overlapping \acp{RIP}. 
Additionally, there is the opportunity to determine the depth by finding a transformation rule from pixel intensity to depth by fitting a 5th order ratio-polynomial model \cite{Laurenzis2009}. 
Optionally, the mapping between pixel intensity and depth can be learned by neural networks \cite{Gruber2018} or regression trees \cite{Adam2017}. 
This enables the flexible adaption of gated settings to any scene to get an optimal image. 
The aforementioned methods estimate the depth pixel-wise, meaning the range of non-illuminated and saturated pixels cannot be determined correctly. 
Therefore, an image-based method has been developed that exploits a convolutional neural network to utilize semantic context across gated images \cite{Gruber2019}. 
However, the existing framework does not provide any measurement to confirm the exact estimation of non-illuminated and saturated pixels. 
For this reason, aleatoric uncertainty is introduced in this work to determine the correctness of the network's output.

\begin{figure*}[ht]
	\centering
	\resizebox{\textwidth}{!}{
	\input{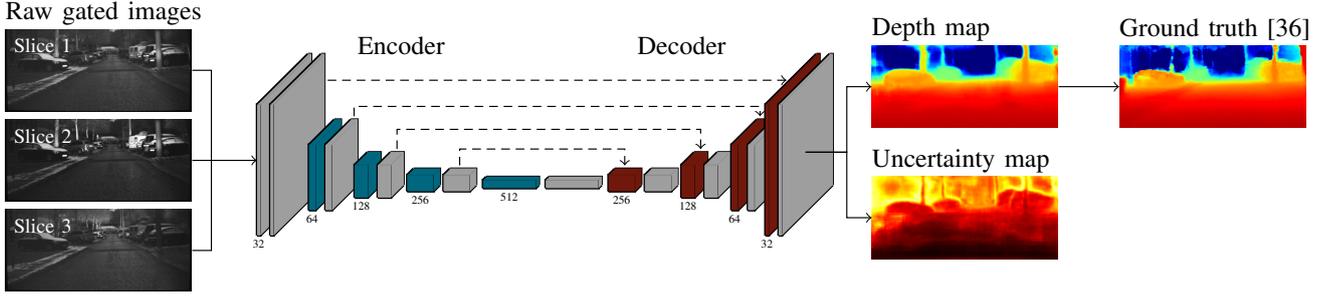}
	}
	\caption{We extend the \emph{Gated2Depth} architecture~\cite{Gruber2019} with an uncertainty output and train the network based on image-guided interpolated \ac{LiDAR} ground truth \cite{ma2018self}.}
	\label{fig:architecture}
	\vspace*{-5mm}
\end{figure*}

%%%%%%%%%%%%%%%%%%%%%%%%%%%%%%%%%%%%%%%%%%%%%%%%%%%%%%%%%%%%%%%%%%%%%%%%%%%%%%%%
\section{METHOD}

\subsection{Gated depth estimation} 

A gated viewing system consists of a diffused flash illuminator and a synchronized gated image sensor enabling high-contrast imaging in low-light, at night, and in adverse weather conditions~\cite{bijelic2018benchmarking}.
Suppose that the laser pulse $p_i(t)$ is reflected by a dominating lambertian reflector with albedo $\alpha$ at distance $r$ and the camera gate $g_i(t)$ is delayed by $\xi_i$.
Then, the \ac{RIP} $C_i(r)$ for each gated setting $i \in \left\{ 1,2,3 \right\}$ is defined by the correlation of $p_i$ and $g_i$, namely
\begin{align}
C_i\left( r \right) = \int\limits_{-\infty}^{\infty} g_i\left( t - \xi_i \right) p_i\left(t - \frac{2r}{c}\right) \beta\left( r \right) \textrm{d}t, \label{eq:gating_equation}
\end{align}
where $c$ and $\beta(r)$ denote the speed of light and the distance-dependent atmospheric influence, respectively \cite{Busck2004}.
Note that the \ac{RIP} is defined to be independent of the scene albedo $\alpha$ and thus the final measured intensity $z_{i,uv}$ on the image sensor at pixel position $(u,v)$ is obtained by
\begin{align}
z_{i,uv} = \alpha C_i(r_{uv})  + \eta_p \left( \alpha C_i(r_{uv}) \right) + \eta_g = f_i(r_{uv})
\label{eq:noise_model}.
\end{align}
%and depends on the range $r_{uv}$.
We follow the Poissonian-Gaussian noise model of Foi \textit{et al.}~\cite{Foi2008} with Poissonian photon shot noise $\eta_p$ and Gaussian read-out noise $\eta_g$.

The task of gated depth estimation is to recover the range $r_{uv}$ from multiple gated measurements $\z_{uv} = \left[ z_{1,uv}, z_{2,uv}, z_{3,uv} \right]$ which basically means finding the inverse function $f^{-1}: \mathbb{R}^3 \rightarrow \mathbb{R}$ that minimizes $\left| \rhat_{uv} - r_{uv} \right|$ with 
\begin{align}
\hat{r} = f^{-1}\left( \mathbf{z} \right) = f^{-1}\left( \mathbf{f}(r_{uv}) \right),
\end{align}
where the function $\mathbf{f}(r_{uv}) = \left[ f_1(r_{uv}), f_2(r_{uv}), f_3(r_{uv}) \right]$ describes a vector of modulated noisy gated images that depends on the distance, see Eq.~\eqref{eq:noise_model}.
In previous works, this inverse function $f^{-1}$ has been learned with a fully-connected neural network~\cite{Gruber2018} or a regression tree~\cite{Adam2017} and is applied pixel by pixel.
However, pixel-based gated depth estimation fails in regions with low \ac{SNR}, saturation, shadows, multi-path, and blooming effects because no spatial correlation is exploited.
\emph{Gated2Depth}~\cite{Gruber2019} is a fully convolutional encoder-decoder network that is able to generate full-resolution depth maps from gated images exploiting semantic context to fill these failure regions.
In this work, we rely on the same network architecture from \emph{Gated2Depth} as shown in Fig.~\ref{fig:architecture} and extend this work with a novel uncertainty measure.
The network is a variant of the popular U-net \cite{ronneberger2015u} with skip-connections that consists of an encoder with four pairs of convolutions followed by a max pooling operation, and a decoder with four additional convolutions and transposed convolutions after each pair.
More details can be found in \cite{Gruber2019}.

\subsection{Loss functions}

The main goal of training is to penalize the absolute differences between ground truth depth $r$ and its depth estimate $\hat{r}$.
This can be achieved by the L1 loss
\begin{align}
\mathcal{L}_{\text{L1}}(r, \rhat) = \frac{1}{N} \sum_{u, v} \norm{r_{uv} - \rhat_{uv}},\label{eq:l1_loss}
\end{align}
where $(u,v)$ denotes the pixel position in the image and $N = uv$ the number of pixels.
Nevertheless, the main challenge of training full-image depth regression is the lack of dense ground truth data.
In state-of-the-art automotive datasets such as KITTI, only sparse \ac{LiDAR} measurements are available.
Even accumulating consecutive \ac{LiDAR} point clouds generates depth maps with only 16\,\% coverage \cite{uhrig2017sparsity}.
There exists a variety of approaches that tackle this problem by introducing specific loss functions during training that enforce dense depth output.
Both multi-scale loss $\mathcal{L}_{\text{L1,m}}$ and smoothness loss $\mathcal{L}_{\text{s}}$ enforce the network to generate full-image depth either by upsampling the sparse ground truth for smaller variants of the output ($\mathcal{L}_{\text{L1,m}}$) or by penalizing large depth differences between neighboring pixels ($\mathcal{L}_{\text{s}}$). 
Note that the multi-scale loss $\mathcal{L}_{\text{L1,m}}$ extends and therefore replaces $\mathcal{L}_{\text{L1}}$.
Semantic understanding from synthetic datasets with dense ground truth depth can be transferred to the real-world training by adding an adversarial loss $\mathcal{L}_{\text{adv}}$ based on a frozen synthetic discriminator that has been trained on synthetic dense ground truth \cite{Gruber2019}.
This discriminator penalizes unrealistic looking depth maps.
We exactly follow the formal definitions of $\mathcal{L}_{\text{L1,m}}$, $\mathcal{L}_{\text{s}}$, and $\mathcal{L}_{\text{adv}}$ as described in \cite{Gruber2019}.
The final loss is given by 
\begin{align}
\mathcal{L} = \mathcal{L}_{\text{L1,m}} + \lambda_{\text{adv}} \mathcal{L}_{\text{adv}} + \lambda_{\text{s}} \mathcal{L}_{\text{s}},
\end{align}
where $\lambda_{\text{adv}}$ and $\lambda_{\text{s}}$ denote tunable hyperparameters for weighting the loss components.

\subsection{Learning from densified ground truth}

Since \ac{LiDAR} depth completion methods have shown impressive results in interpolating sparse depth measurements guided by intensity images, we propose a novel training method for \emph{Gated2Depth} that relies on this densified depth as ground truth.
We apply the popular Sparse-to-Dense framework \cite{ma2018self} on the \ac{LiDAR} point clouds and RGB images of the training set.
This certainly limits the performance of \emph{Gated2Depth} to the performance of the depth completion approach.
However, gated depth estimation aims to replace these expensive \ac{LiDAR} systems with cost-sensitive hardware and intelligent post-processing, and achieving depth completion performance is already sufficient.

\subsection{Introducing uncertainty}

Bayesian modeling enables the extraction of epistemic and aleatoric uncertainty of neural networks.
Epistemic uncertainty represents the uncertainty in the model parameters, which results from an insufficient amount of training data and vanishes for an infinite number of data. 
However, aleatoric uncertainty depicts observation noise of the input and remains stable independent of the input quantity.
In \cite{kendall2017uncertainties}, they showed that epistemic uncertainty can identify inputs that deviate from the training dataset, whereas aleatoric uncertainty is appropriate for real-time applications, as no expensive dropout sampling is required. 
Our research will be applied in cars, where fast sensing of the environment is essential. 
Hence, only aleatoric uncertainty is implemented into our framework.

To capture aleatoric uncertainty, the output $r_{uv}$ of the neural network is modeled as a likelihood function. 
For the given depth regression task, we propose a likelihood function $p(r_{uv})$ that follows a Laplacian distribution with mean $\rhat_{uv}$ and variance $\sig_{uv}$:
\begin{equation}
p(r_{uv}) =  \frac{1}{2\sig_{uv}}\exp \left[ -\frac{\norm{r_{uv}-\rhat_{uv}}}{\sig_{uv}} \right].
\end{equation}
Thereby, $\rhat_{uv}$ is the estimated depth for an input $\z_{uv}$ and $\sig_{uv}$ represents the corresponding aleatoric uncertainty. 
Both depth $\hat{r}_{uv}$ and uncertainty $\sig_{uv}$ are modeled as explicit outputs of a single neural network, which is parameterized by its model weights.
Instead of a Laplacian distribution, a Gaussian distribution can be utilized as a likelihood function. 
However, this has led to worse results in our regression problem and follows the experiences that a L1 loss is more effective for depth regression~\cite{carvalho2018regression}.
To find the model parameters that explain the model best, the likelihood function has to be maximized, which corresponds to the minimization of the negative log-likelihood, given by
\begin{equation}
-\log p(r_{uv}) = \frac{||r_{uv}-\rhat_{uv}||}{\sig_{uv}} + \log \sig_{uv} + \log 2. \label{eq:negative_log_likelihood}
\end{equation}
Therefore, the negative log-likelihood averaged over each pixel $(u,v)$ is considered as aleatoric loss function for training the neural network, namely
\begin{equation}
\mathcal{L}_{\text{L1,aleatoric}} = \frac{1}{N} \sum_{u,v} \norm{r_{uv}-\rhat_{uv}} e^{-s_{uv}} + s_{uv},
\end{equation}
where $N$ is the number of pixels of an output image and $s_{uv} = \log \sig_{uv}$ is the log variance that is numerically more stable as it avoids division by zero. 
Additionally, the constant term $\log 2$ in Eq.\eqref{eq:negative_log_likelihood} is neglected. 
When aleatoric uncertainty is applied on multiple scaled versions of the output, the multi-scale aleatoric loss $\mathcal{L}_{\text{L1,aleatoric,m}}$ is given by
\begin{align}
\mathcal{L}_{\text{L1,aleatoric,m}} = \sum_{i=0}^{M-1} \lambda_{m_i} \mathcal{L}_{\text{L1,aleatoric}}(r^{(i)}, \hat{r}^{(i)}, \hat{s}^{(i)}), 
\end{align}
where $r^{(i)}$, $\hat{r}^{(i)}$, and $\hat{s}^{(i)}$ are scaled versions of $r$, $\hat{r}$, and $\hat{s}$.
To train and test models without aleatoric uncertainty, $s_{uv}$ has to be set to zero and the aleatoric loss $\mathcal{L}_{\text{L1,aleatoric,m}}$ converges to the multi-scale L1 loss $\mathcal{L}_{\text{L1,m}}$.

In conclusion, to introduce aleatoric uncertainty into the \emph{Gated2Depth} architecture, we add an uncertainty map as second output and train the whole network with an aleatoric loss $\mathcal{L}_{\text{L1,aleatoric,m}}$ that replaces the multi-scale loss $\mathcal{L}_{\text{L1,m}}$. 
Finally, the full loss is obtained by
\begin{align}
\mathcal{L} = \mathcal{L}_{\text{L1,aleatoric,m}} + \lambda_{\text{adv}} \mathcal{L}_{\text{adv}} + \lambda_{\text{s}} \mathcal{L}_{\text{s}}.
\end{align}

\subsection{Uncertainty filtering}

Gated depth estimation provides depth information at pixel level.
However, low \ac{SNR} or saturated pixels can infer depth only from context which is not always possible.
The uncertainty estimation provides a measure of how confident the model is about the depth estimation.
To show the benefit of our proposed confidence measure, we introduce \emph{uncertainty filtering} where depth estimates are filtered out when the uncertainty value is above a threshold $t$.
This results in an overall better performance because obviously wrong estimates with high uncertainty are neglected.
As a baseline, we use a simple \ac{SNR} filtering based on the laser illumination, assuming that depth estimation from pixels with low illumination variance is insufficient.
Given a set of gated input pixels $\mathbf{z}_{uv} = [z_{1,uv}, z_{2,uv}, z_{3,uv}]$ and a predefined threshold $\vartheta$, pixels with low illumination variance are defined as the ones that satisfy $\mathrm{max}(\mathbf{z}_{uv})-\mathrm{min}(\mathbf{z}_{uv})<\vartheta$.

%%%%%%%%%%%%%%%%%%%%%%%%%%%%%%%%%%%%%%%%%%%%%%%%%%%%%%%%%%%%%%%%%%%%%%%%%%%%%%%%
\section{DATASETS}

Since gated imaging is an emerging sensor technology, there are not many state-of-the-art datasets that provide gated images.
Gruber \textit{et al.}~\cite{Gruber2019} have presented the first long-range gated dataset in real-world automotive scenarios that consists of 14,277 samples recorded in Northern Europe.
They have equipped a vehicle with a Brightway Vision BrightEye gated camera with two \ac{VSCEL} illuminators at \unit[808]{nm} in the bumper.
In addition to gated images (1920\,$\times$\,1024, \unit[10]{bit}), stereo images (1920\,$\times$\,1024, \unit[12]{bit}) and \ac{LiDAR} point clouds (64 lines) from a Velodyne HDL64-S3 laser scanner are provided.
For experiments with full-resolution ground truth depth, we rely on 9,804 simulated gated images based on the \ac{GTA V} computer game \cite{Gruber2019}.
We follow the same training, validation and test splits as in \emph{Gated2Depth}~\cite{Gruber2019} in order to make our contribution comparable.

%%%%%%%%%%%%%%%%%%%%%%%%%%%%%%%%%%%%%%%%%%%%%%%%%%%%%%%%%%%%%%%%%%%%%%%%%%%%%%%%
\section{Assessment}

\subsection{Experimental setup}

Besides the implementation of uncertainty, our framework provides further changes compared to the original \textit{Gated2Depth} network \cite{Gruber2019}. 
To enlarge the field-of-view of the input images, the original crop of 150 pixels at each side is replaced by a simple upper crop of 152 pixels, which was necessary due to missing \ac{LiDAR} measurements in this area.  
Furthermore, instead of training two separate models for day and night, a single daytime-independent model is built, which simplifies the subsequent application in cars. 

The presented method is implemented in \emph{tensorflow} and trained with an \emph{adam} optimizer and a learning rate of 0.0001. 
We have empirically chosen $\lambda_{\text{s}}$ and $\lambda_{\text{adv}}$ to 10.
All models utilized here are trained for 15 epochs on a GeForce GTX TITAN Xp \ac{GPU}, which took roughly 13 hours for the synthetic and 20 hours for the real dataset. 
According to the validation \ac{MAE}, we have selected the best performing epoch and hyperparameters. 
For the synthetic dataset the error is evaluated in a range from \unit[3-150]{m} and for the real dataset from \unit[3-80]{m} due to the limited range of the applied \ac{LiDAR} system. 
We rely on the popular metrics \ac{RMSE}, \ac{MAE}, \ac{SIlog} and the thresholds $\delta_i < 1.25^i$
for $i \in \{1, 2, 3\}$ as defined in \cite{Gruber2019}.

\subsection{Results on simulated data}

\begin{figure}[t]
\centering
\setlength{\tabcolsep}{1mm}
\resizebox{\columnwidth}{!}{
\begin{tabular}{@{}ccc@{}}
Full gated & Depth & Uncertainty \\
\includegraphics[width=0.4\columnwidth]{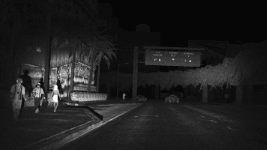} &
\includegraphics[width=0.4\columnwidth]{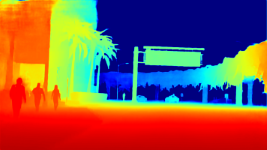} &
\includegraphics[width=0.4\columnwidth]{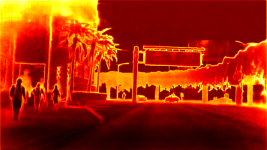} \\
\includegraphics[width=0.4\columnwidth]{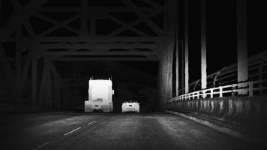} &
\includegraphics[width=0.4\columnwidth]{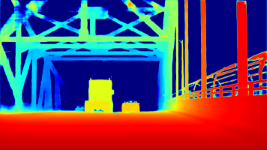} &
\includegraphics[width=0.4\columnwidth]{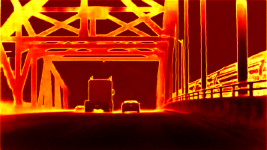} \\
\includegraphics[width=0.4\columnwidth]{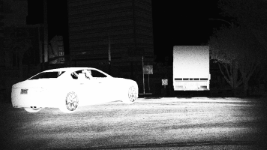} &
\includegraphics[width=0.4\columnwidth]{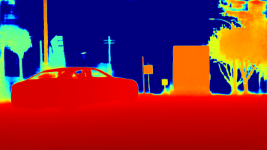} &
\includegraphics[width=0.4\columnwidth]{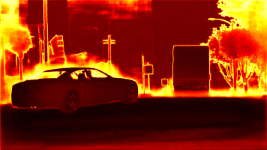} \\
\end{tabular}}
\vspace*{-2mm}
\caption{Qualitative examples for the synthetic dataset. For color coding, we refer to Fig.~\ref{fig:teaser}.}
\label{fig:qualitative_example_syn}
\vspace*{-5mm}
\end{figure}

\begin{figure}[t]
	\centering
	\resizebox{\columnwidth}{!}{
		\begin{tikzpicture}
						\begin{axis}
							[
							xlabel=Coverage {[\% of all pixels]},
							ylabel=MAE {[m]},
							xmin=17, xmax=100,
							ymin=0, ymax=4,
							legend style={
								cells={anchor=west},
								legend pos=north east,
								font=\footnotesize
							},
							legend columns=3, 
							legend entries={SNR filter, Uncertainty filter},
							height=4cm,
							width=\linewidth
							]
					
						\addlegendimage{very thick, dai_deepred}
						\addlegendimage{very thick, dai_petrol}					

						\addplot+ 	[
									thick,
									rot,
									no marks,
									]
						table[x index=0,y index=1,col sep=space]{fig/mae_coverage_syn/snr.txt};
						
						\addplot+ 	[
															thick,
															mittelblau,
															no marks,
															]
						table[x index=0,y index=1,col sep=space]{fig/mae_coverage_syn/aleatoric.txt};

						\end{axis}
			\end{tikzpicture}
	}
	\vspace*{-6mm}
    \caption{Accuracy measured in MAE with respect to the depth coverage for \emph{synthetic data}. By increasing the filter thresholds, unreliable depth estimations are filtered out resulting in a better overall performance.}
    \label{fig:mae_coverage_syn}
    \vspace*{-2mm}
\end{figure}
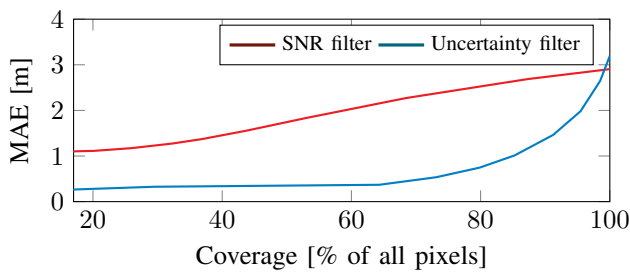

For the synthetic dataset, the conventional and the aleatoric model are trained from scratch without adversarial and smoothness loss, because dense ground truth is available. 
Qualitative examples for the model with uncertainty are illustrated in Fig.~\ref{fig:qualitative_example_syn}. 
The bright areas in the uncertainty maps indicate high uncertainty and can be perceived especially at contours of objects and in non-illuminated areas of the image. 
Fig.~\ref{fig:mae_coverage_syn} compares \ac{SNR} and uncertainty filtering based on the estimated uncertainty.
The performance of the aleatoric model can be significantly improved compared to the conventional one by filtering only a small number of pixels. 
To reduce the \ac{MAE} by half, only about 10\,\% of the pixels of the aleatoric model must be filtered whereas the conventional model requires filtering of about 60\,\% to achieve such an error reduction.
Note that the conventional model is slightly better than the one with uncertainty which explains the different starting points at 100\,\% coverage.

\subsection{Ablation study for loss and ground truth}

\begin{figure}[t]
	\centering
	\setlength{\tabcolsep}{1mm}
	\resizebox{\columnwidth}{!}{
		\begin{tabular}{ccc}
			1 epoch & 3 epochs & 10 epochs \\
			MAE = \unit[3.33]{m} & MAE = \unit[3.04]{m} & MAE = \unit[2.53]{m} \\
			\includegraphics[width=0.33\columnwidth]{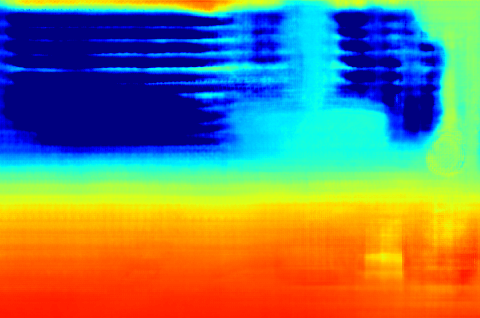} &
			\includegraphics[width=0.33\columnwidth]{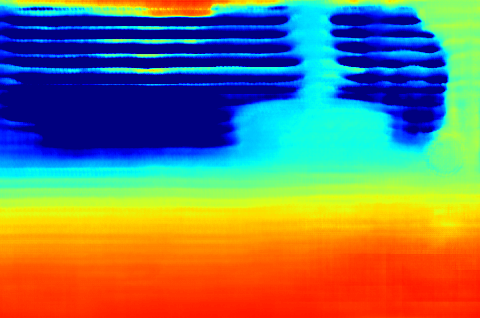} &
			\includegraphics[width=0.33\columnwidth]{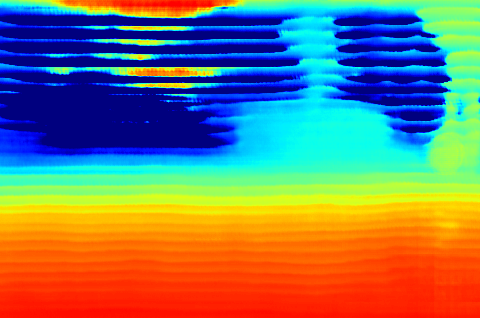} \\
	\end{tabular}}
	\vspace*{-2mm}
	\caption{This figure shows how training on sparse \ac{LiDAR} ground truth depth generates horizontal patterns for an increasing number of epochs, although the MAE on the validation set decreases.}
	\label{fig:stripes}
	\vspace*{-2mm}
\end{figure}

\begin{table}[t]
\caption{Ablation study for different loss functions and ground truth data types trained on real data.}
\resizebox{\columnwidth}{!}{
\setlength{\tabcolsep}{1mm}
\setlength{\extrarowheight}{2pt}

\begin{tabular}{@{}ccccc@{}}
\toprule
\multirow{2}{*}{\textbf{Ground truth}} & \multirow{2}{*}{\textbf{Loss} $\mathcal{L}$}  & \multicolumn{3}{c}{\textbf{MAE [m]}} \\ 
 & & \ac{LiDAR} & DC & \ac{LiDAR}+DC \\ \midrule
\ac{LiDAR} & $\mathcal{L}_{\text{L1}}$ & 2.57 &  3.66 & 3.12\\
\ac{LiDAR} & $\mathcal{L}_{\text{L1,m}}$ & 2.66 &  3.74 & 3.20\\
\ac{LiDAR} & $\mathcal{L}_{\text{L1}} + \lambda_{\text{s}}\mathcal{L}_{\text{s}}$ & 2.98 & 3.31  & 3.15\\
\ac{LiDAR} & $\mathcal{L}_{\text{L1}} + \lambda_{\text{adv}}\mathcal{L}_{\text{adv}}$ & 2.56 & 3.88  &  3.22\\
\ac{LiDAR} & $\mathcal{L}_{\text{L1,m}} + \lambda_{\text{s}}  \mathcal{L}_{\text{s}} +
\lambda_{\text{adv}}  \mathcal{L}_{\text{adv}}$ & 2.85 & 3.17 & \textbf{3.01} \\ \midrule
DC & $\mathcal{L}_{\text{L1}}$ & 3.08 & 2.64 & \textbf{2.86}\\
DC & $\mathcal{L}_{\text{L1,m}} + \lambda_{\text{s}}  \mathcal{L}_{\text{s}} +
\lambda_{\text{adv}}  \mathcal{L}_{\text{adv}}$ & 3.25 & 2.81 & 3.03 \\
\bottomrule
\end{tabular}}
\label{fig:ablation_study}
\vspace*{-4mm}
\end{table}

\begin{figure}[t]
\centering
\setlength{\tabcolsep}{1mm}
\resizebox{\columnwidth}{!}{
\begin{tabular}{@{}ccc@{}}
{\footnotesize \ac{LiDAR}} & {\footnotesize \ac{LiDAR}} & {\footnotesize Depth completion} \\[-1mm]
{\scriptsize $\mathcal{L}_{\text{L1}}$} & {\scriptsize$\mathcal{L}_{\text{L1,m}} + \lambda_{\text{s}}  \mathcal{L}_{\text{s}} +
\lambda_{\text{adv}}  \mathcal{L}_{\text{adv}}$} & {\scriptsize$\mathcal{L}_{\text{L1}}$} \\
\includegraphics[width=0.33\columnwidth]{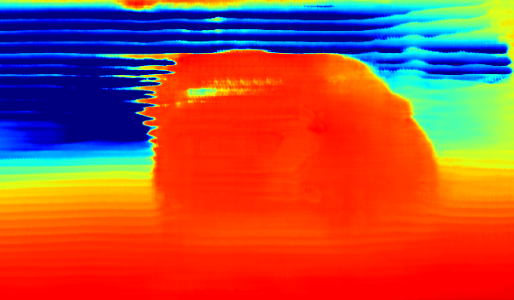} &
\includegraphics[width=0.33\columnwidth]{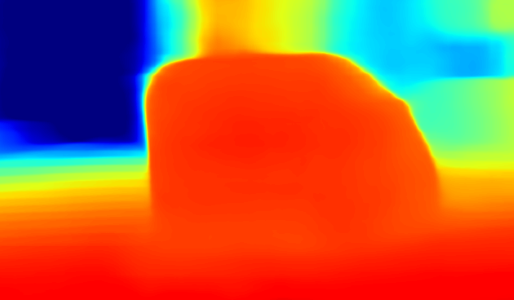} &
\includegraphics[width=0.33\columnwidth]{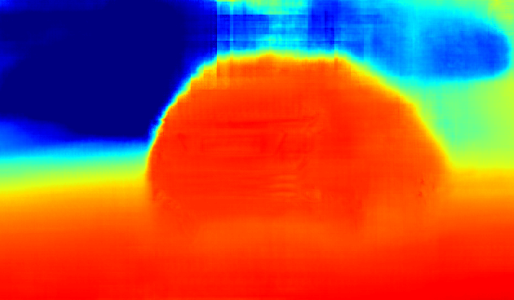} \\
\end{tabular}}
\vspace*{-2mm}
\caption{Horizontal stripe patterns that occur when training on sparse ground truth \ac{LiDAR} can be removed by additional loss components or extended ground truth annotations. 
}
\label{fig:stripe_removal}
\vspace*{-3mm}
\end{figure}

When training on real data with sparse \ac{LiDAR} ground truth without any countermeasures, horizontal patterns in the estimated depth maps occur as Fig.~\ref{fig:stripes} illustrates.
For an increasing number of epochs, these horizontal patterns get even worse, although the \ac{MAE} decreases due to evaluation on the sparse \ac{LiDAR} points only.
To quantify the horizontal patterns, we additionally evaluate on the \ac{DC} ground truth.
This dense ground truth is obtained by Sparse-to-Dense \cite{ma2018self}, a \ac{LiDAR} depth completion method based on RGB images and sparse \ac{LiDAR} points.
While the error on the \ac{LiDAR} ground truth reflects the accuracy of the depth estimation, the \ac{DC} error measures the strength of the horizontal patterns.

Multi-scale loss, smooth loss, and adversarial loss are well-known approaches to handle sparse ground truth depth.
Additionally, we propose a training on depth completed ground truth as another countermeasure.
We evaluated different loss combinations for \ac{LiDAR} and \ac{DC} ground truth after training for 10 epochs with synthetic model initialization. 
To ensure accurate and smooth depth maps, the average of \ac{LiDAR} and \ac{DC} error is computed to compare the different approaches. 
The ablation study in Tab.~\ref{fig:ablation_study} shows that applying a multi-scale loss with additional smoothness and adversarial loss delivers the best performance for models trained on sparse \ac{LiDAR} ground truth.
When trained on \ac{DC} ground truth, a simple L1 loss generates the best result and additional loss components do not help. 
Fig.~\ref{fig:stripe_removal} illustrates the significant reduction of the horizontal patterns by using a multi-loss function (multi-scale, smooth, adversarial) or depth completed ground truth.
According to the average of \ac{LiDAR} and \ac{DC} \ac{MAE} metric, the model trained on \ac{DC} ground truth shows a slightly better performance.
Moreover, models with dense ground truth can be trained for longer, since no horizontal patterns are generated.

\subsection{Results on real data}

\begin{figure}[t]
\centering
\setlength{\tabcolsep}{1mm}
\resizebox{\columnwidth}{!}{
\begin{tabular}{@{}ccc@{}}
Full gated & Depth & Uncertainty \\
\includegraphics[width=0.4\columnwidth]{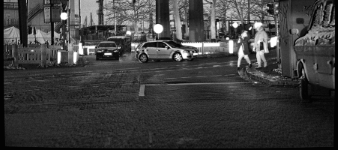} &
\includegraphics[width=0.4\columnwidth]{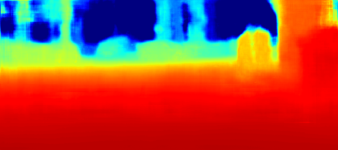} &
\includegraphics[width=0.4\columnwidth]{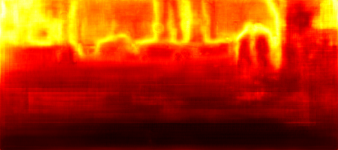} \\
\includegraphics[width=0.4\columnwidth]{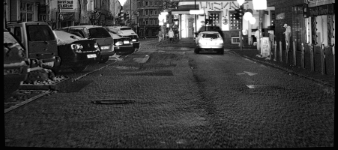} &
\includegraphics[width=0.4\columnwidth]{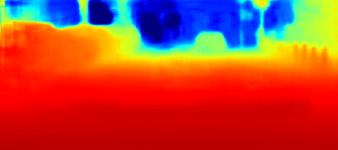} &
\includegraphics[width=0.4\columnwidth]{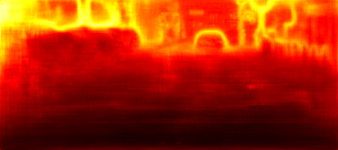} \\
\includegraphics[width=0.4\columnwidth]{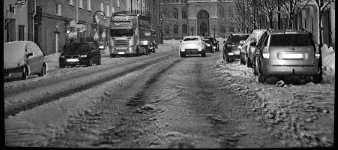} &
\includegraphics[width=0.4\columnwidth]{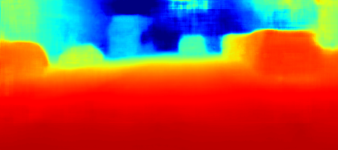} &
\includegraphics[width=0.4\columnwidth]{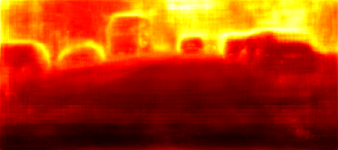} \\
\end{tabular}}
\vspace*{-2mm}
\caption{Qualitative examples for the real dataset. In particular, high uncertainty arises at object edges and shadows above the objects.}
\vspace*{-6mm}
\label{fig:qualitative_example_real}
\end{figure}

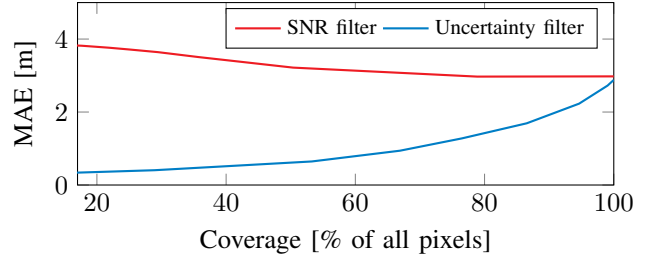
\begin{figure}[t]
	\centering
	\vspace{.1em}
	\resizebox{\columnwidth}{!}{
		\begin{tikzpicture}
						\begin{axis}
							[
							xlabel=Coverage {[\% of all pixels]},
							ylabel=MAE {[m]},
							xmin=17, xmax=100,
							ymin=0, ymax=5,
							legend style={
								cells={anchor=west},
								legend pos=north east,
								font=\footnotesize
							},
							legend columns=2, 
							height=4cm,
							width=\linewidth
							]
						
						\addplot+ 	[
								thick,
								rot,
								no marks,
								]
						table[x index=0,y index=1,col sep=space]{fig/mae_coverage_real/snr.txt};
						\addlegendentry{SNR filter}

						\addplot+ 	[
								thick,
								mittelblau,
								no marks,
								]
						table[x index=0,y index=1,col sep=space]{fig/mae_coverage_real/aleatoric.txt};
						\addlegendentry{Uncertainty filter}			

						\end{axis}
			\end{tikzpicture}
	}
	\vspace*{-6mm}
    \caption{Accuracy measured in MAE with respect to the depth coverage for real data. We create the points by increasing the filter threshold.}
    \label{fig:mae_coverage_real}
    \vspace*{-3mm}
\end{figure}

\begin{figure}[ht!]
\centering
\setlength{\tabcolsep}{1mm}
\resizebox{\columnwidth}{!}{
\begin{tabular}{@{}ccc@{}}
\textbf{100\,\% coverage} & \textbf{90\,\% coverage} & \textbf{80\,\% coverage} \\
MAE = \unit[3.32]{m} & MAE = \unit[2.14]{m} & MAE = \unit[1.50]{m} \\
\includegraphics[width=0.4\columnwidth]{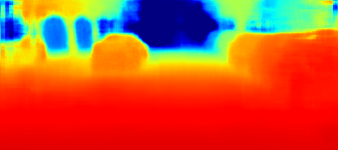} &
\includegraphics[width=0.4\columnwidth]{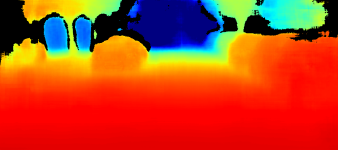} &
\includegraphics[width=0.4\columnwidth]{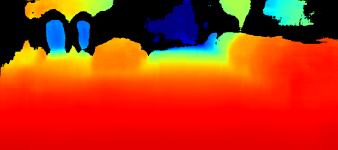}  
\\[1mm]
\textbf{70\,\% coverage} & \textbf{60\,\% coverage} & \textbf{50\,\% coverage} \\
MAE = \unit[1.11]{m} & MAE = \unit[0.72]{m} & MAE = \unit[0.51]{m} \\
\includegraphics[width=0.4\columnwidth]{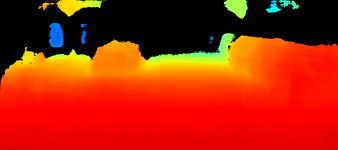} &
\includegraphics[width=0.4\columnwidth]{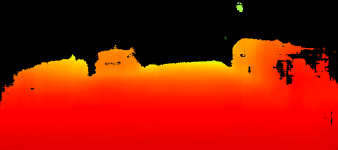} &
\includegraphics[width=0.4\columnwidth]{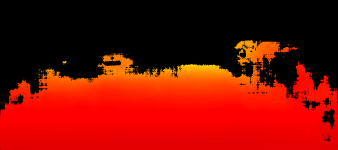} \\
\end{tabular}}
\vspace*{-2mm}
\caption{Uncertainty filtering for varying coverage.}
\label{fig:uncertainty_filtering}
\vspace*{-7mm}
\end{figure}

\begin{figure*}[t]
    \hspace{-8pt}\resizebox{1.03\linewidth}{!}{
    {
    \setlength{\tabcolsep}{1mm} 
    \begin{tabular}{cccccc}
    	RGB & 
    	Full gated & 
    	\ac{LiDAR} &  
    	\monodepthLegend & 
    	\sgmLegend &
    	[m] \\

    	\includegraphics[width=0.25\textwidth]{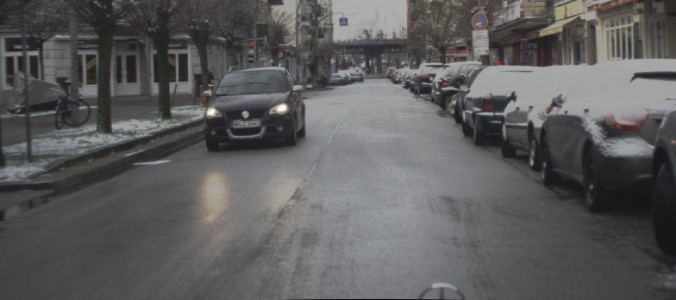} &
    	\includegraphics[width=0.25\textwidth]{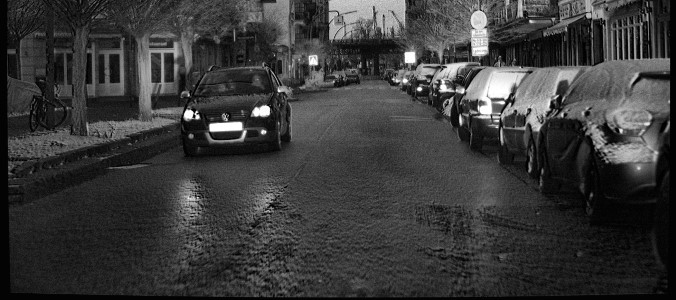} &
    	\includegraphics[width=0.25\textwidth]{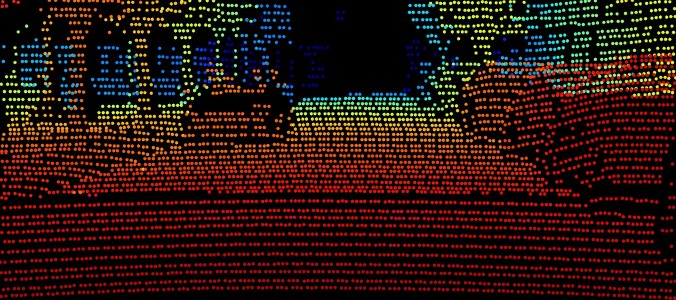} &
    	\includegraphics[width=0.25\textwidth]{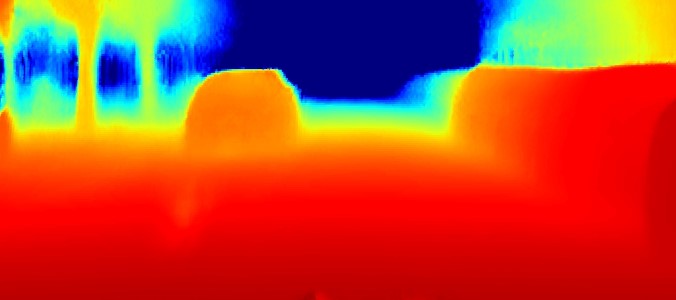} &
    	\includegraphics[width=0.25\textwidth]{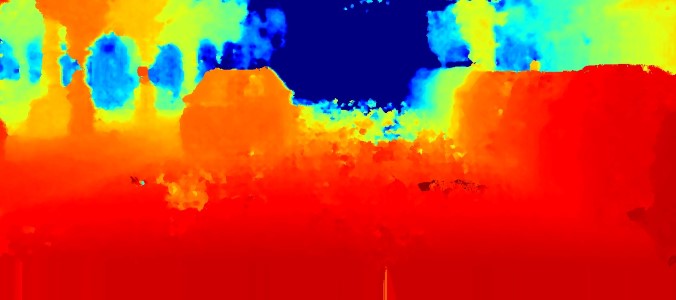} &
    	\multirow{3}{*}[1.7cm]{\hspace*{-2mm}\includegraphics[height=4.5cm]{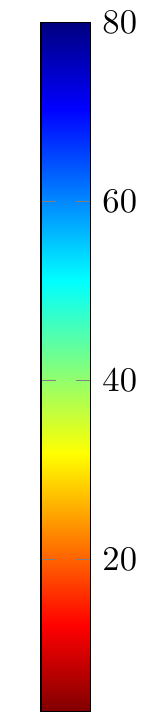}}
    	\\
    	
    	Gated2Depth \cite{Gruber2019} &
    	G2D+ Uncertainty &
    	G2D+ Filter (80\,\%) &
    	\sparseLegend &
    	\psmLegend &
    	\\
    	\includegraphics[width=0.25\textwidth]{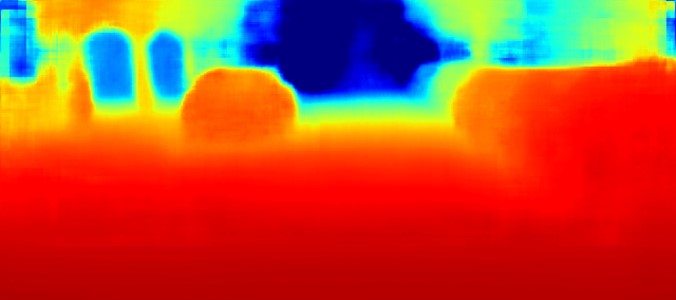} &
    	\includegraphics[width=0.25\textwidth]{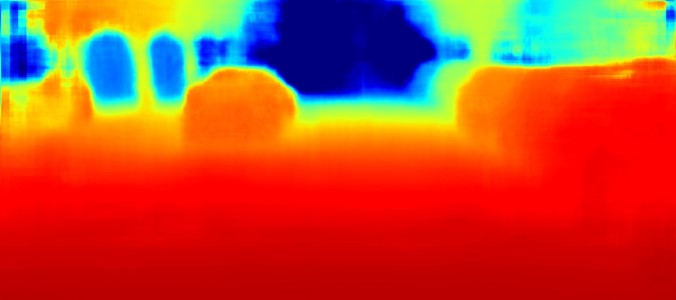} &
    	\includegraphics[width=0.25\textwidth]{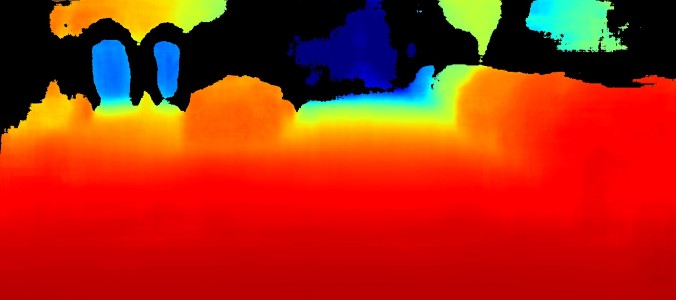} &
    	\includegraphics[width=0.25\textwidth]{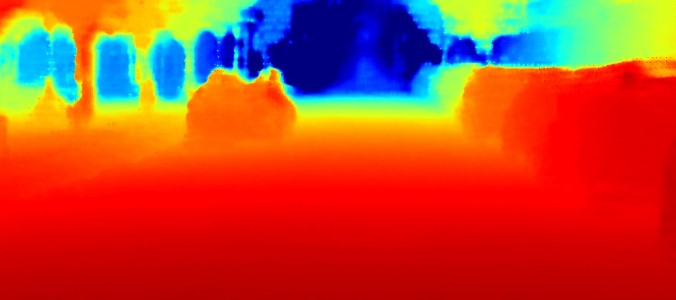} &
    	\includegraphics[width=0.25\textwidth]{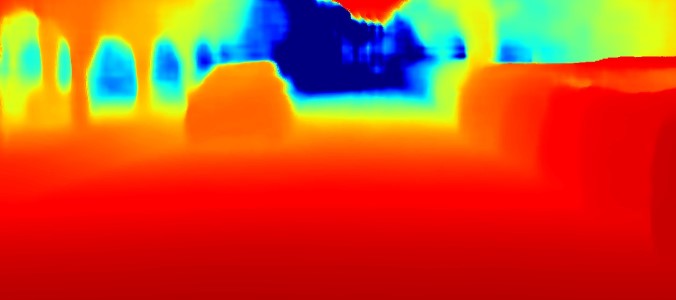} &

        \\
        
    \end{tabular}
    }
    }
    \vspace*{-2mm}
    \caption{Experimental daytime results for a variety of state-of-the-art methods.}
    \label{fig:qualitative_results_real}
    \vspace*{-5mm}
\end{figure*}

\begin{table}[t]
	\caption{Comparison of our proposed framework \emph{G2D+} and state-of-the-art methods on the real test dataset according to common metrics as utilized in \cite{gruber2019pixel}. Note that Sparse-to-Dense requires ground truth input.}
	\centering
	\hspace*{-5mm}
	\resizebox{\columnwidth}{!}{
		\setlength{\tabcolsep}{1mm} 
		\setlength{\extrarowheight}{2pt}
		\begin{tabular}{@{}l|ccccccc@{}}
			\toprule
			\multirow{2}{*}{\textbf{Method}} & \textbf{RMSE} & \textbf{MAE} & \textbf{SIlog} &  $\boldsymbol{\delta_1}$ &  $\boldsymbol{\delta_2}$ &  $\boldsymbol{\delta_3}$ & \textbf{Compl.} \\
			& [m] & [m] & & $\left[ \% \right]$  & $\left[ \% \right]$ & $\left[ \% \right]$ & \\
			\midrule
			\textsc{\monodepthLegend} &9.59 & 4.70 & 25.35 & 82.32 & 92.23 & 95.68 & 1.00 
 \\
			\textsc{\psmLegend} &6.71 & 2.45 & 19.35 & 92.65 & 95.85 & 97.27 & 1.00 
 \\
			\textsc{\sgmLegend} &10.83 & 5.26 & 37.82 & 77.38 & 86.36 & 90.52 & 1.00 
 \\
			\textsc{\sparseLegend} &5.77 & 1.64 & 18.39 & 94.91 & 96.38 & 97.39 & 1.00 
 \\
			\textsc{Gated2Depth \cite{Gruber2019} } &7.08 & 2.98 & 21.78 & 89.86 & 94.91 & 96.76 & 1.00 
 \\
			\midrule
			\textsc{G2D+ Uncertainty} &7.00 & 2.88 & 21.22 & 90.28 & 94.92 & 96.80 & 1.00 
 \\
			\textsc{G2D+ Filter (80\,\%)} & 3.84 & 1.49 & 13.78 & 95.53 & 98.06 & 98.75 & 0.87 
 \\
			\bottomrule
	\end{tabular}}
	\label{tab:quantitative_results}
\end{table}

\begin{figure}[t]
	\centering
	\vspace*{-3mm}
	\resizebox{\linewidth}{!}{
	\radarChart{./fig/results/radar_chart_uncertainty_paper_test_all.txt}}
	\vspace*{-6mm}
	\caption{Kiviat diagram that visualizes the results from Table~\ref{tab:quantitative_results}. We normalize each metric such that the best approach is at 1 and the worst approach at 0.1.}
	\label{fig:radar_chart}
	\vspace*{-5mm}
\end{figure}

We train a model with uncertainty and one without uncertainty (baseline) to investigate how the incorporation of uncertainty changes the overall performance.
Qualitative examples for our proposed uncertainty model in Fig.~\ref{fig:qualitative_example_real} exhibit high uncertainty for object contours and non-illuminated areas, e.g. due to shadows above each object.
Tab.~\ref{tab:quantitative_results} indicates that the introduction of uncertainty comes at no additional costs as there is no loss in performance.
The model with uncertainty has even a slightly better performance than the baseline model without uncertainty. 
The benefit of our proposed uncertainty measure is shown by comparing \ac{SNR} and uncertainty filtering. 
As Fig.~\ref{fig:mae_coverage_real} shows, \ac{SNR} filtering does not help to get rid of erroneous measurements. 
On the contrary, \ac{SNR} filtering probably removes pixels with good depth estimates and therefore decreases the overall performance.
By applying the novel uncertainty filtering, the \ac{MAE}
can be significantly lowered. 
To halve the \ac{MAE}, only about 20\,\% of the pixels have to be filtered out.
Fig.~\ref{fig:uncertainty_filtering} demonstrates an exemplary filtering process for different pixel coverages. 
The \ac{MAE} can be reduced from \unit[3.32]{m} to less than \unit[1]{m} at 50\,\% coverage. 
However, when filtering out too many pixels, only the foreground is preserved and thus the semantic context of the image disappears. 
Hence, we decided that obtaining about 80\,\% of the image pixels is a good choice since the \ac{MAE} is reduced by half and individual objects can still be extracted.

\subsection{Comparison with state-of-the-art methods}

We follow \cite{Gruber2019} and compare our extended Gated2Depth method (\emph{G2D+}) with state-of-the-art methods for 3D environment perception, such as monocular depth estimation \cite{Godard2017}, stereo vision \cite{hirschmuller2005accurate,Chang2018} and \ac{LiDAR} depth completion \cite{ma2018self}.
\emph{Monodepth} \cite{Godard2017} and \emph{PSMnet} \cite{Chang2018} are finetuned on the real gated dataset.
Finetuning of \emph{Sparse-to-Dense} \cite{ma2018self} is not possible because neither dense nor semi-dense ground truth depth is available.
While Tab.~\ref{tab:quantitative_results} provides the evaluation metrics on the whole test dataset (day+night), the radar chart in Fig.~\ref{fig:radar_chart} visualizes the results in a normalized representation. 
Each metric is normalized to the range $[ 0,1 ]$ such that the best approach is at 1 and the worst approach at 0.1. 
The results clearly show that traditional stereo \cite{hirschmuller2005accurate} and monocular depth estimation \cite{Godard2017} show worst performance, while \ac{LiDAR} depth completion \cite{ma2018self} provides the best results.
However, \ac{LiDAR} depth completion relies on sparse ground truth depth input from \ac{LiDAR} and it is hard for other approaches to achieve this performance.
Note that compared to \emph{Gated2Depth} in \cite{Gruber2019}, we do not use separate models for day and night and we evaluate on a significantly larger image crop.
Gated depth estimation without any filter shows in this setting only similar performance as deep stereo \cite{Chang2018}.
However, the additional uncertainty maps can be used to filter out unreliable depth estimates and \emph{Gated2Depth} with uncertainty filter even outperforms depth completion performance.

%%%%%%%%%%%%%%%%%%%%%%%%%%%%%%%%%%%%%%%%%%%%%%%%%%%%%%%%%%%%%%%%%%%%%%%%%%%%%%%%
\section{CONCLUSIONS}

This work extends the recent \emph{Gated2Depth} method with aleatoric uncertainty that provides additional confidence information for each depth estimate.
We show in the application of uncertainty filtering, how this uncertainty measure can help to filter out unreliable depth estimates increasing the overall system performance.
In an ablation study, we show that our proposed training on RGB guided interpolated ground truth depth is superior to conventional multi-loss approaches.
Exciting future research includes the application of uncertainty maps into sensor fusion approaches, either for object detection or scene understanding enabling the integration of gated viewing systems into recent sensor setups of safe self-driving cars.

%%%%%%%%%%%%%%%%%%%%%%%%%%%%%%%%%%%%%%%%%%%%%%%%%%%%%%%%%%%%%%%%%%%%%%%%%%%%%%%%

\vspace*{2mm}
\noindent\footnotesize{This work has received funding from the European Union under the H2020 ECSEL Programme as part of the DENSE project, contract number 692449.}

%%%%%%%%%%%%%%%%%%%%%%%%%%%%%%%%%%%%%%%%%%%%%%%%%%%%%%%%%%%%%%%%%%%%%%%%%%%%%%%

\bibliographystyle{IEEEtran}
\bibliography{ref}

\end{document}